\documentstyle[12pt]{article}
\def\be{\nopagebreak[3]\begin{equation}}
\newcommand{\ee}{\end{equation}}
\def\begar{\begin{array}}
\def\ea{\end{array}}

\newcommand{\scs}[1]{{\scriptscriptstyle #1}}

\newcommand{\eq}[1]{Eq.~(\ref{#1})}

\newcommand{\qru}{q^{n}\! = \! 1}
\renewcommand{\d}{\partial}
\renewcommand{\ss}[1]{{\hbox{}_{#1}}}
\newcommand{\D}[3]{D_{\scs{#1,}}\hbox{}^{#2}_{#3}}
\newcommand{\barD}[3]{\overline{D}_{\scs{#1,}}\hbox{}^{#2}_{#3}}

\newcommand{\R}{\scs R}
\newcommand{\ba}{\bar{a}}
\newcommand{\bb}{\bar{b}}
\newcommand{\bc}{\bar{c}}
\newcommand{\bd}{\bar{d}}

\begin{document}
\begin{titlepage}
\begin{flushright}
NBI-HE-93-27\\
June 1992
\end{flushright}
\vspace*{36pt}
\begin{center}
{\huge \bf
qQCD{\large \bf 2} and G/G model}
\end{center}
\vspace{2pc}
\begin{center}
 {\Large D.V. Boulatov}\\
\vspace{1pc}
{\em The Niels Bohr Institute\\
University of Copenhagen\\
Blegdamsvej 17, 2100 Copenhagen \O \\
Denmark}
\vspace{2pc}
\end{center}
\begin{center}
{\large\bf Abstract}
\end{center}
The 2D lattice gauge theory with a quantum gauge group $SL_q(2)$ is
considered. When $q=e^{i\frac{2\pi}{k+2}}$, its weak coupling
partition function
coincides with the one of the G/G coset model ({\em i.e.} equals the
Verlinde numbers). However, despite such a remarkable coincidence, these
models are not equivalent but, in some certain sense, dual to each
other.

\vfill
\end{titlepage}

\newpage

The necessity of a regularization in quantum field theories naturally leads
to their lattice formulation. In the context of gauge theory, lattice
models \cite{Wilson} play a rather fundamental role of its
non-perturbative definition. Within the strong coupling expansion,
lattice QCD gives some evidence for the confinement phenomenon.
In topological theories, the lattice formulation is extremely useful
since it establishes a direct connection with invariants of different
types of complexes.
For example, in order to calculate the partition function in a
topological theory, we can, in principle, consider the simplest cell
decomposition of a base manifold. An implementation of such ideology
was given in ref.~\cite{Boul} (see also \cite{Boul1}),
where the notion of q-deformed 3D lattice gauge
theory ($qQCQ_3$) was introduced and shown to be connected with the
Turaev-Viro (and, hence, Chern-Simons) invariant.

The aim of the present paper is to apply the general approach of
ref.~\cite{Boul} to the 2D lattice gauge theory with a compact quantum
gauge group ($qQCD_2$). As will be shown, for the
$SL_q(N)$, $\qru$, gauge group, its weak coupling limit gives
a well defined topological gauge model on an arbitrary cell
decomposition of a 2D manifold. In the case $|q|<1$, we obtain
$q$-analogs of all classical (\hbox{$q=1$}) quantities without changing the
nature of the model qualitatively. $QCD_2$ itself has recently drawn
attention because of the new insights brought about by the Gross-Taylor
reinterpretation of its partition function as the sum over branched
coverings  \cite{Gross}. The topological nature of $QCD_2$ was noticed
long ago by A.A.Migdal \cite{Mig} who proposed to use a character expansion of
Boltzman weights to manifest it. Following this guideline, the partition
function of $QCD_2$ on an arbitrary 2-manifold was obtained in the form of
a sum over irreps of a gauge group in ref.~\cite{Rus1}. Recently, on
the sphere, the third order phase transition with respect to the area
was found \cite{DougKaz} (see also \cite{Rus2}, where the partition
function for small areas was calculated). Wilson loop averages were
investigated in refs.~\cite{KazKos}. In the $qQCD_2$ case, one
should distinguish between over and under-crossings of Wilson lines
obtaining in this way a connection with the knot invariants.

However, $QCD_2$ is not the only 2D theory possessing a gauge symmetry. Another
well-known example is the gauged WZW model \cite{Gaw}. In
ref.~\cite{Witt,Spieg}
the topological G/G coset model was suggested, which was investigated
then from various viewpoints in refs.~\cite{Yank}. Its partition
function is given by the number of conformal blocks in the
WZW model on a corresponding manifold \cite{Witt}. These numbers were
calculated by E.Verlinde in ref.~\cite{Verl}. In the present paper, we
show that the quantum group structure of the Verlinde numbers
naturally appears in the framework of lattice gauge theory. However,
amplitudes in the G/G model have no natural representation in $qQCD_2$,
and so do observables in the latter. We shall argue that these two
models are actually dual to each other.

\bigskip

The partition function in $qQCD_2$ can be defined in close analogy
with the standard (classical) lattice gauge model \cite{Wilson,Mig}. Let
$\cal L$ be an arbitrary 2D lattice, {\em i.e.} a cell decomposition of
a closed oriented 2D manifold such that every link $\sigma^1_i$ divides
exactly two 2D cells: $\delta\sigma^1_i=\sigma^2_n-\sigma^2_m$. Its
end points $(\sigma^0_k,\sigma^0_\ell)$ can be also written down
as the formal difference:
$\d\sigma^1_i=\sigma^0_k-\sigma^0_\ell$ (the signs are agreed with an
orientation). By definition the boundary of a 2-cell
$\sigma^2_k$ is the formal ordered sum of 1-cells entering in it:
$\d\sigma^2_k=\sum\sigma^1_i$. We attach a variable,
$g_i$, taking values in a gauge group to every link, $\sigma^1_i$,
so that the change of an orientation, $\sigma^1_i\to-\sigma^1_i$,
corresponds to the conjugation: $g_i\to g_i^{\scs -1}$. Gauge variables
attached to different links are mutually commutative while at the same
link not. Taking the ordered product of gauge variables along the
boundary of a 2-cell $\sigma^2_i$, we define the holonomy

\be
h_i=\prod_{\d\sigma^2_i}g_k
\ee
The Boltzmann weights are positive-definite functions of the holonomies
which can be defined by the character expansion

\be
W(h_i)=\sum_\R d_\R\chi_\ss{R}(h_i)w_\R(e^\scs{2})
\label{bolwei}
\ee
where the Fourier coefficients $w_\R(e^\scs{2})$ are functions of the gauge
coupling constant $e^\scs{2}$; $d_\R$ is the quantum dimension of an $R$'th
irrep. If for $e^\scs{2}=0$ all $w_\R(0)=1$,
the r.h.s. of \eq{bolwei} becomes the group
$\delta$-function. We shall call this choice of the weights the weak
coupling limit. As characters are central and every gauge variable
enters in exactly two different characters, there is no problem with
non-commutativity and we can define the $qQCD_2$ partition function
simply by multiplying the Boltzmann weights and integrating over all
field configurations

\be
Z({\cal L})=\int\prod_{\sigma^1_i}dg_i\prod_{\sigma^2_j}
W(\prod_{\d\sigma^2_j}g_k)
\label{parfun}
\ee
The integrals in this equation extract multiplicities of the trivial
representation in tensor products appearing after substituting the
Fourier expansion (\ref{bolwei}) for $W(\prod_{\d\sigma^2_j}g_k)$. Let
$\D{R}{a}{b}(g)$ denote a matrix element of an irrep $R$
($a,b=1,\ldots,dim(R)$ are representation indices). Then, by definition,
for all non-trivial representations ({\em i.e.} $\D{R}{a}{b}(g)\neq 1$),

\be
\int dg\ \D{R}{a}{b}(g) = 0
\ee

If $|q|<1$, representations of $SU_q(N)$ repeat the ones of the
classical group $SU(N)$. The standard normalization in this case is
$\int dg\ 1 = 1$. However, in our case,
we need to truncate representations in
\eq{bolwei} to the fusion ring\footnote{Actually, we have already done it by
introducing the quantum dimensions in the definition of the Boltzmann
weights} (in the standard notation for $SL_q(2)$, it means
$J=0,\frac{1}{2},1,\frac{3}{2},\ldots,\frac{k}{2}$).
As all integrations are reduced to the orthogonality of matrix
elements, indecomposable representations will never appear. Having in
mind analogy with finite groups, let us normalize the integral in
this case as

\be
\int dg\ 1 = \sum_{\R}d_\R^2 \equiv |G_q|
\label{norm}
\ee
For $SL_q(2)$, this number (the quantum rank) is well known:

\be
|SL_q(2)|=\frac{k+2}{2\sin^2\frac{\pi}{k+2}}
\ee

\bigskip
In what follows, we shall need a machinery to work with quantum matrix
elements, etc. Let us now give the set of rules and definitions we shall
use. It will be convenient to assume sums over all repeated indices (the
Einstein summation rule).
The conjugation in the quantum case is ambiguous, therefore let us
use the bar to denote conjugate indices:

\be
\D{R}{a}{b}(g^{\scs-1})=\barD{R}{\bb}{\ba}(g)
\ee
and, in principle,
they can be matched with the initial ones in many different ways.
In ribbon Hopf algebras there exists at least one invertible element
$C^{\ba}_{b}$
which does it \cite{ReshTur}. Its square changes a framing by 1 ({\em
e.g.} in the $SL_q(2)$ case \cite{KiResh},
$C^{a}_{\bc}C^{\bc}_{b}=q^{2J(J+1)}\delta^a_b$ and
$\overline{C}^{a}_{\bc}\overline{C}^{\bc}_{b}=q^{-2J(J+1)}\delta^a_b$,
where $\overline{C}^{a}_{\bc}$ is the
inverse $\overline{C}^{a}_{\bc}C^{\bc}_{b}=\delta^a_b$).
We shall not
use it in $qQCD_2$ and, hence, shall not take care of framings. Instead,
we shall use the conjugation matrix

\be
\Lambda_{a\bb}=\sqrt{d_\R}\langle RaRc|0\rangle C^{c}_{\bb}\hspace{1cm}
\Lambda^{a\bb}=\sqrt{d_\R}\langle 0|R\bc R\bb\rangle
\overline{C}^{a}_{\bc}
\ee
where $\langle RaSb|Tc\rangle$ is the Clebsch-Gordan coefficient ({\em
i.e.} matrix elements of an irrep $T$
in the decomposition of the tensor product $V_R\otimes V_S$).
We also introduce the inverses

\be
\Lambda_{a\bb}\Lambda^{\bb c}=\delta^{c}_{a}
\hspace{2cm}
\Lambda^{a \bb}\Lambda_{\bb c}=\delta^{a}_{c}
\ee
Now, we can unambiguously identify

\be
\D{R}{b}{a}(g^{\scs -1})=\Lambda_{a\bc}\barD{R}{\bc}{\bd}(g) \Lambda^{\bd b}
\ee
and the unitarity condition reads

\be
\D{R}{b}{a}(g)\Lambda_{b\bc}\barD{R}{\bc}{\bd}(g)\Lambda^{\bd e} =
\delta^{e}_{a}
\ee

We define the quantum character

\be
\chi_{\ss{R}}(g)=
\D{R}{a}{b}(g)\Lambda_{a\bc}\Lambda^{b\bc}
\ee
which is a cyclic function

\be\begar{l}
\chi_{\ss{R}}(gh)=
\D{R}{a}{d}(g)\D{R}{d}{b}(h)\Lambda_{a\bc}\Lambda^{b\bc}=
\D{R}{d}{b}(h)\Lambda_{d\bc}\barD{R}{\bc}{\ba}(g^{\scs -1})\Lambda^{b\ba}=\\
\D{R}{d}{b}(h)\D{R}{b}{a}(g)\Lambda_{d\bc}\Lambda^{a\bc}=
\chi_{\ss{R}}(hg)
\ea
\ee
The dimensions are the characters of the unity

\be
d_\R=\chi_{\ss{R}}(I)=\Lambda_{a\bc}\Lambda^{a\bc}
\ee

The orthogonality condition for matrix elements reads

\be
\int dg\ \D{R}{a}{b}(g)\barD{S}{\bc}{\bd}(g) =
\frac{|G_q|}{d_\R}\delta_{\scs{RS}}\Lambda^{a\bc}\Lambda_{b\bd}
\ee

The set of objects and operations just introduced
is sufficient for our purposes,
however, it should be stressed that it would not be such in the $qQCD_3$
case, for example. Say, we have not defined the R-matrix and
indeed we shall not need it in what follows.

The $SL_q(2)$ gauge group will be our main example. In this case we
identify

\be
\ba=-a\hspace{1cm} -J\leq a\leq J
\ee
then

\be
\Lambda_{a\bb}=(-1)^a q^{\frac{a}{2}}\delta_{a+b,0} \hspace{.5cm}
\Lambda^{a\bb}=(-1)^a q^{\frac{a}{2}}\delta_{a+b,0} \hspace{1cm}
d_{\scs{J}}=\frac{\sin\Big(\frac{2J+1}{k+2}\pi\Big)}
{\sin\Big(\frac{\displaystyle\pi}{k+2}\Big)}
\ee

To obtain the topological model we have to put the Boltzman weights to
be the $\delta$-functions. With our normalization (\ref{norm}), we have
to take in \eq{bolwei} $w_\R(0)$ independent of an irrep $R$ and such
that $\int dg\ W(g)=1$, {\em i.e.}, $w_{\scs{J}}(0)=1/|G_q|$.
As we shall see, with this choice, the partition function (\ref{parfun})
will take only integer values exactly coinciding with the Verlinde
numbers. To calculate it we have to use the orthogonality condition
subsequently

\be
\int dg\ \chi_{\ss{R}}(g^{\scs -1}h)\chi_{\ss{S}}(gf)
=\frac{|G_q|}{d_\R}\delta_{\scs{RS}}\chi_{\ss{R}}(hf)
\ee
(which can be interpreted as the unification of 2-cells)
until we get only one 2-cell. For a genus $p$ Riemann surface,
${\cal M}_p$, we find

\be
Z({\cal M}_p)=\int\prod_{i=1}^{p}dg_i dh_i\
W\Big(\prod_{i=1}^{p}g_ih_ig^{\scs-1}_ih^{\scs-1}_i\Big)=
\sum_{\R} \Big(d^2_\R/ |G_q|)^{1-p}
\label{Z=}
\ee
For $SL_q(2)$, this formula takes the familiar form \cite{Verl}
\be
Z({\cal M}_p)=\sum_{J=0,\frac{1}{2},...}^{k/2}
\Bigg(\frac{2\sin^2\Big(\frac{2J+1}{k+2}\pi\Big)}{k+2}\Bigg)^{1-p}
\ee
If $p=0$ (the sphere), we have no integrations and the answer is just
$Z(S^2)=W(I)=1$.

The equation (\ref{Z=}) has the following natural
interpretation. The number of gauge variables is equal to the number of
generators of the fundamental group $\pi_1({\cal M}_p)$,
and the argument of the
$\delta$-function under the integral is its defining relation. Hence,
the integral counts the number of inequivalent representations of
$\pi_1({\cal M}_p)$ by elements of the gauge group.

\bigskip
The partition function of the G/G topological coset
model on a Riemann surface of a
genus $p$ has exactly the same form (\ref{Z=}), which suggests that
these two models has to be closely related. To understand this relation,
let us consider amplitudes, which in the G/G model coincide with
the fusion rules in the corresponding WZW model. On the other hand,
the complete set of observables in gauge theory is given by Wilson loop
operators. In topological case, we can obtain non-trivial answers only
for non-contractible loops, {\em i.e.} encircling punctures and/or
handles. An arbitrary Riemann surface can be obtained by sewing
spheres with three punctures. Having fixed holonomies around the
punctures, we can write

\be
Z(S^2_{ijk})=\frac{1}{|G_q|}\sum_\R\frac{1}{d_\R}\chi_{\ss{R}}(h_i)
\chi_{\ss{R}}(h_j)\chi_{\ss{R}}(h_k)
\label{gauge_amp}
\ee

In the G/G model, we fix representation indices at punctures and
the amplitude has the form of a fusion rule which can be written as the
integral

\be
N_{R_iR_jR_k}=\frac{1}{|G_q|}\int dg\ \chi_{\ss{R_i}}(g)
\chi_{\ss{R_j}}(g)\chi_{\ss{R_k}}(g)
\label{wzw_amp}
\ee
If we took a finite group, Eqs.~(\ref{gauge_amp}) and (\ref{wzw_amp})
would be exactly dual to each other. The latter could be considered as a
gauge model in which representation indices play a role of gauge variables.
In general, there is no exact duality between
representations and group elements, however, as we see, the topological
$qQCD_2$ and the G/G model are formally dual to each other.

Let us finish with the remark that, for knotted but otherwise contractible
loops, Wilson averages will coincide with corresponding Jones polynomials.
Because the loops intersect only at lattice sites, we have to insert the
R-matrices at corresponding points to construct Wilson loop operators
correctly.

It would be interesting to establish a connection of our approach with
the one of Ref.~\cite{FR}. Despite some obvious similarity, it is not yet
clear whether they are really equivalent.

\bigskip
{\large\bf Acknowledgments}

\medskip
I would like to thank A.Alekseev, A.N.Kirillov, M.Petropoulos
and V.Sadov for the discussions.

\end{document}